\documentstyle[12pt,aaspp4]{article}
 
\def\hi{\noindent \hangindent=2.5em}
\begin{document}
\title{The Fueling of Nuclear Activity:
II. The Bar Properties
of Seyfert and Normal Galaxies}
\author{John S. Mulchaey\altaffilmark{1,2}}
\affil{The Observatories of the Carnegie Institution of Washington,
813 Santa Barbara St., Pasadena, CA 91101}

\centerline{and}

\author{Michael W. Regan\altaffilmark{3}}
\affil{Department of Astronomy, University of Maryland, College Park, MD 20742}

\altaffiltext{1}{Visiting Astronomer, Kitt Peak National Observatory,
National Optical Astronomy Observatories, operated by the Association of
Universities for Research in Astronomy, Inc., under contract with the National
Science Foundation.}
\altaffiltext{2}{mulchaey@pegasus.ociw.edu}
\altaffiltext{3}{mregan@astro.umd.edu}

\begin{abstract}

We use a recent near-infrared imaging survey of samples of Seyfert
and normal galaxies to study the role of bars in the fueling of 
nuclear activity.
The active galaxy sample includes Seyfert galaxies in the 
Revised Shapely-Ames (RSA) and Sandage \& Tammann's (1987) 
extension to this catalog. The normal galaxies were selected to
match the Seyfert sample in Hubble type, redshift,
inclination and blue luminosity. 
All the galaxies in both samples classified as barred in the RSA
catalog are also barred in the near-infrared. In addition, $\sim$
55\% of the galaxies classified as non-barred in the RSA show evidence for
bars at 2.1 $\mu$m. 
Overall, $\sim$ 70\% of the galaxies observed show evidence for bar 
structures.
The incidence of bars in the Seyfert and normal 
galaxies is similar, suggesting Seyfert nuclei do not occur preferentially
in barred systems. Furthermore, 
a slightly higher percentage of 
normal galaxies have multiple-bar structures.

A significant percentage of the Seyfert galaxies in our sample show no evidence for
the presence of a bar even in the near-infrared. This suggests that
either large-scale kiloparsec bars are not a universal fueling mechanism
in Seyfert galaxies
or that the bars in the non-barred Seyferts were recently destroyed,
possibly by
the formation of the central black hole.
\end{abstract}

\section{\bf Introduction}

Galactic bars are frequently invoked as candidates for facilitating
the transfer of mass from the interstellar medium of Seyfert galaxies to their
central engines (e.g. Schwartz 1981; Norman 1987; Shlosman, Frank
 \& Begelman 1989).
However, the importance of bars 
in Seyfert galaxies remains controversial, with some
studies suggesting that Seyfert nuclei occur preferentially in 
barred systems (e.g. Arsenault 1989), while others find no such preference
(Heckman 1980; Simkin, Su, \& Schwarz 1980;
Moles et al. 1995). Most comparisons of the hosts of normal and 
active galaxies have been made at optical wavelengths, where the
effects of dust and star formation can mask bar structures.
 Near-infrared images 
provide a better tracer of the mass distribution including bars since 
the relative importance of these effects is reduced 
at near-infrared wavelengths.
Several K-band imaging studies of 
Seyfert galaxies have revealed
 the presence of bar structures in galaxies classified
as unbarred in the optical (e.g. Scoville et al. 1988; Thronson et al. 1989;
McLeod \& Reike 1995), demonstrating the benefit of working in the near-infrared.
In this {\it Letter}, we use a recently completed modified 
K-band imaging survey of samples of Seyfert and normal galaxies
(Mulchaey, Regan \& Kundu 1997; hereafter Paper I) to study the role
of bars in the fueling of nuclear activity.

\section{\bf Sample Selection and Bar Identification}

The Seyfert sample was selected from the Revised Shapely-Ames (RSA) and 
the Sandage \& Tammann (1987) extension to the RSA. All of the galaxies observed
have recessional velocities less than 5000 km s$^{-1}$ and logarithmic 
axial ratios (Log a/b) less than 0.2.
The axial ratio limit was adopted
to avoid highly inclined galaxies where
bars can be difficult to recognize. The control sample of normal galaxies
was also selected from the RSA and was matched to the Seyfert sample
in Hubble type, redshift, inclination and blue luminosity. 
A total of 30 Seyfert and 25 normal galaxies were observed in the near-infrared
with a modified K-band filter. A complete description of the sample
properties and observing conditions are given in Paper I.

 The bar morphology
of each galaxy was determined using the criteria outlined in detail 
in Paper I. 
Approximately 50\% of the galaxies in both samples are classified as barred in
the RSA catalog. We detect near-infrared bars in all of these galaxies. 
We also find evidence for bars in sixteen of the twenty-nine galaxies classified
as unbarred in the RSA. 
This further demonstrates the gains that can be made 
by observing galaxies in the near-infrared. Overall, $\sim$ 70\% of
the galaxies observed show evidence for a bar at 2.1$\mu$m.

\section{\bf The Incidence of Bars in Seyfert and Normal Galaxies}

Figure 1 summarizes the bar statistics in Paper I.
We note that the percentage of barred systems is comparable
in the Seyfert and normal galaxy samples.
This suggests that Seyfert nuclei
 do not occur preferentially in barred galaxies,  a result consistent
with earlier studies at optical wavelengths 
 (Heckman 1980; Simkin, Su, \& Schwarz 1980;
Moles et al. 1995). A slightly higher percentage of
the Seyfert 2 galaxies than Seyfert 1 galaxies appear
to be barred in our images (83\% vs. 67\%, respectively), but this result is not
statistically significant given the small number of objects of each type.

While we find no significant difference between
 the percentage of barred galaxies in our normal
and active galaxy samples, there
 may still be differences between the Seyfert hosts
and spiral galaxies in general. For example, in selecting our control sample,
we have matched the normal galaxies 
to the Seyfert galaxies in total blue luminosity.
However, the Seyfert host galaxies tend to be more luminous on average than the 
typical spiral galaxy in the RSA catalog. Thus, our control galaxies are more 
luminous in the blue than typical RSA spirals and our results for the control sample
may not reflect the characteristics of typical spirals. In fact, we note that the 
percentage of bars in both the Seyfert and control sample is higher in the RSA
(i.e. 42\% and 45\%) than the overall percentage of barred galaxies in the RSA 
for similar Hubble types (
for example, only 25\% of the RSA spirals of type Sa or Sab are 
barred). Therefore there may be a trend between the luminosity of the host and the 
presence of a bar. While ideally we would like to make a direct comparison between
the bar properties of our Seyfert galaxies and the RSA spirals in general, 
near-infrared images for large samples of spirals over a range in luminosities do
not yet exist. From the present data, we can conclude that the percentage of 
barred galaxies is comparable among Seyfert galaxies and normal spiral galaxies of
similar luminosity.

\section{Bars As A Fueling Mechanism}

Shlosman, Frank \& Begelman (1989) have proposed a
\lq\lq bars within bars\rq\rq \ mechanism that would drive material
collected from a large-scale bar
down to the scales where the
central supermassive blackhole dominates the gravitational potential
($\sim$ inner 10 pc). In their model, gas driven inward from the large-scale
stellar bar accumulates in the central few hundred parsecs in a rapidly
rotating disk. If the mass of the gas in this disk is an appreciable
fraction ($>$ 20$\%$) of the dynamical mass at that radius, the disk
may become unstable and a gaseous bar can form. A dynamically unstable
series of such bars may exist, continuing the inflow of material towards the nucleus.

Nested bars have been observed in some galaxies (e.g.
Shaw et al. 1995; Wozniak et al. 1995;
Friedli et al. 1996), possibly supporting the above scenario.
In fact, Wozniak et al. (1995) have noted a high percentage of Seyfert galaxies
in their sample of double-barred systems.
 We have 
searched for the presence of multiple bar structures in our near-infrared 
images. Figure 1
shows, however,  that the percentage of double-barred galaxies is 
actually higher in our normal galaxy sample than in our Seyfert sample.
The failure to find multiple bars in most of
 our galaxies could be a result of our 
limited resolution. At the typical distance of our sample galaxies, our spatial
resolution is $\sim$ 300--500 pc (H$_{\rm o}$ = 50 km s$^{-1}$ Mpc$^{-1}$ assumed). 
Higher resolution images may
uncover many more examples of multiple bars. We also note that the 
\lq\lq bars within bars\rq\rq \ mechanism proposed by Shlosman et al. (1989)
relies on {\it gaseous} bars in the inner kiloparsec and these bars may not
always be accompanied by the stellar bars that our images probe. High 
resolution observations of atomic and molecular gas in Seyferts may be 
required to fully test the Shlosman et al. (1989) scenario.

Piner, Stone \& Teuben (1995)
have recently suggested that bars with a high axial ratio
(i.e. thin bars) can also move mass efficiently into the inner 100 pc 
of a galaxy. A comparison of the maximum ellipticity for the
bars in the two samples (Figure 2) suggests no
differences in the bar axial ratios of the 
Seyfert and normal galaxies.
Furthermore, the two galaxies with the highest bar axial ratios
are normal galaxies. 
Thus, we find
 no evidence that Seyfert galaxies preferentially
contain thin bars.

\section{Non-Barred Seyferts}

While bars appear to be a common feature in Seyfert galaxies,
 $\sim$ 30\% of our sample
galaxies show no evidence for a bar even in the near-infrared.  
It seems likely that most of these galaxies do not contain
bars. A particularly striking example is the Seyfert 1 galaxy
NGC 7213, which has nearly circular isophotes from the center down to a
surface brightness level of $\sim$ 19 K mag arcsecond$^{-2}$.
The failure to find bars in even this small percentage of 
Seyferts indicates that bars are probably not a ubiquitous
feature of active galaxy hosts. Mcleod \& Reike (1995) reached
a similar conclusion from their K-band study of the CfA Seyfert sample. 

A relevant question is whether the non-barred Seyferts show evidence
for other perturbations that might be responsible for fueling the
nuclear activity. Galaxy interactions and encounters have often 
been suggested as a mechanism to induce activity in galaxies
and in fact several of the non-barred Seyferts appear to
be interacting with neighboring galaxies (e.g. NGC 5427; Kennicutt 
\& Keel 1984).
However, in other cases, the nearest neighbor
is much too far away for gravitational interactions to be a 
reasonable scenario (e.g. NGC 788 nearest neighbor is 
over 1 h$_{\rm 50}$ Mpc away; Moles et al. 1995). In addition,
 when a major merger interaction drives gas 
toward the center of a
galaxy it does so by forming a bar (Mihos \& Hernquist 1996).
Since these possibly interacting Seyferts are non-barred, 
it is not clear how the interaction could be transporting mass to the 
central engine.

Moles et al. (1995) have
suggested that although bars may not be 
more common in Seyfert galaxies than 
normal spirals, other features often associated with bars, such as rings,
may be more prevalent in galaxies with nuclear activity.
However, rings are a secondary probe of the potential and
not as straightforward a probe as our near-infrared images. Thus,
a non-detection of a bar in the near-infrared probably rules out the 
presence of a strong nonaxisymmetric component to the potential
even when the galaxy has a ring.

\section{Destruction of Bars in Active Galaxies}

It has been proposed that the formation of a black hole
at the center of a galaxy
can lead to the destruction of a bar (Norman et al. 1996; 
Friedli \& Benz 1993).
The simulations of Norman et al. (1996) show that the bar can be
destroyed in only a fraction of the bar rotation time 
once the mass of the central black hole approaches 5\% of the
total mass of the galaxy.
For most barred galaxies this means that bars can be destroyed 
in only $\sim$ 10$^7$ years.
In this case, 
the stars that made up the bar are distributed into random axisymmetric
orbits that give the appearance of a bulge.
This very fast destruction time scale means that it may be possible for
a black hole at the center of a galaxy to remain active even if the
bar that supplied the fuel has been destroyed.

These non-barred Seyferts should exhibit certain characteristics if they are
formerly barred galaxies.
The stellar population of the bulge of these galaxies should be the
same age as the stellar population of the disk.
In addition, the central velocity dispersion of these galaxies should
be quite high since the central black hole should contain $\sim$5\%
of the total mass of the galaxy.
Therefore, the velocity dispersion of the centers of the non-barred
Seyferts should imply a higher mass fraction in the central region than
the velocity dispersions of the central regions of the 
barred Seyfert galaxies, where the ratio of black hole to
galaxy mass is presumably lower.

\section{Conclusions}

We have used a large near-infrared imaging survey to study the 
bar properties of Seyfert and 
normal galaxies. We find near-infrared bars in all of the sample galaxies
previously classified as barred in the Revised Shapely-Ames catalog and
in $\sim$ 55\% of the galaxies previously classified as unbarred.  
Approximately 70\% of the Seyfert galaxies are barred, with 
$\sim$ 10\% having multiple-bar structures. The percentage of 
bars is comparable in the normal galaxy sample.
In general,
the global properties of bars,
such as axial ratio, appear to be similar in Seyfert and normal spiral
galaxies of similar luminosity.

A significant fraction of the Seyferts studied show no evidence for
bars even in the near-infrared. This suggests that either large-scale
bars are not a universal mechanism for the transfer of mass to 
the central engine
or that the bars in the observed non-barred Seyferts were destroyed,
possibly
with the formation of the central black hole.

The authors acknowledge useful discussions with Alice Quellin and 
valuable comments from the anonymous referee.
JSM acknowledges support from a Carnegie postdoctoral fellowship.

\vfill\eject

\centerline{References}

\hi{Arsenault, R. 1989, AA, 217, 66}

\hi{Friedli, D. \& Benz, W. 1993, AA, 268, 65}

\hi{Friedli, D., Wozniak, H., Rieke, M., Martinet, L. \& Bratschi, P. 1996, AA Supl, 118, 461}

\hi{Heckman, T. M. 1980, AA, 87, 142}

\hi{Kennicutt, R. C. \& Keel, W. C. 1984, \apj, 279, L5}

\hi{McLeod, K. K., \& Reike, G. H. 1995, \apj, 441, 96}

\hi{Mihos, J. C., \& Hernquist, L. 1996, ApJ, 464, 641}

\hi{Moles, M., Marquez, I., \& Perez, E. 1995, \apj, 438, 604}

\hi{Mulchaey, J. S., Regan, M. W., \& Kundu, A. 1997, \apjs, in press}

\hi{Norman, C. 1987, in {\it Galactic and Extragalactic Star Formation},
(eds. Pudritz, R. E., \& Fich, M.), Kluwer, Dordrecht, p495}

\hi{Norman, C. A., Sellwood, J. A., \& Hasan, H. 1996, \apj, 462, 114}

\hi{Piner, B. G., Stone, J., \& Teuben, P. 1995, ApJ, 449, 508}

\hi{Sandage, A., \& Tammann, G. A. 1981, {\it A Revised Shapely-Ames Catalog of Bright
Galaxies},(Carnegie Institution of Washington, Washington, D. C.),
 Publ. 635, 1st edition. (RSA)}

\hi{Sandage, A., \& Tammann, G. A. 1987, {\it A Revised Shapely-Ames Catalog of Bright
Galaxies}, (Carnegie Institution of Washington, Washington, D. C.), Publ. 635, 2nd edition.}

\hi{Schwartz, M. 1981, \apj, 247, 77}

\hi{Scoville, N. et al. 1988, \apj, 311, L47}

\hi{Shaw, M., Axon, D., Probst, R., \& Gatley, I. 1995, MNRAS, 274, 369}

\hi{Shlosman, I., Begelman, M. C., \& Frank, J. 1990,
Nature, 345, 679}

\hi{Simkin, S. M., Su, H. J., \& Schwarz, M. P. 1980, \apj, 
237, 404}

\hi{Thronson, H. A. et al. 1989, \apj, 343, 158}

\hi{Wozniak, H., Friedli, D., Martinet, L., Martin, P., \&
Bratschi, P. 1995, AA Supl, 111, 115}

\vfill\eject

\centerline{Figure Captions}

\noindent{Figure 1--Fraction of barred and double-barred systems in the Seyfert and 
control galaxy samples. The percentage of barred galaxies is comparable 
in the Seyfert and control samples. A slightly higher percentage of the normal
galaxies have double bars.}

\vskip 0.2cm

\noindent{Figure 2--Histogram of the maximum ellipticities for the bars in  
the Seyfert and control galaxy samples. There is no evidence that
the bars in Seyferts are preferentially thin (i.e. high ellipticity).}




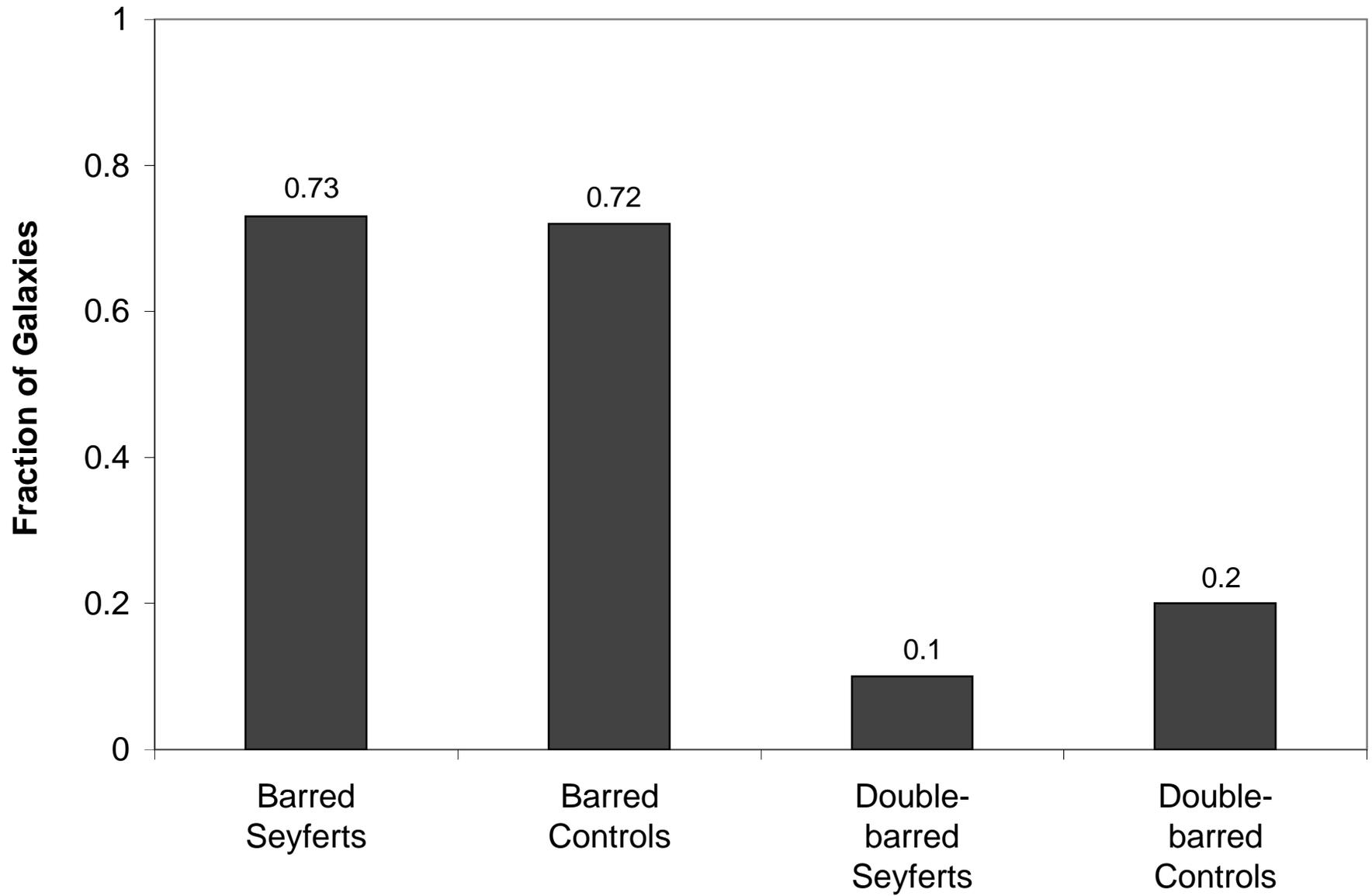
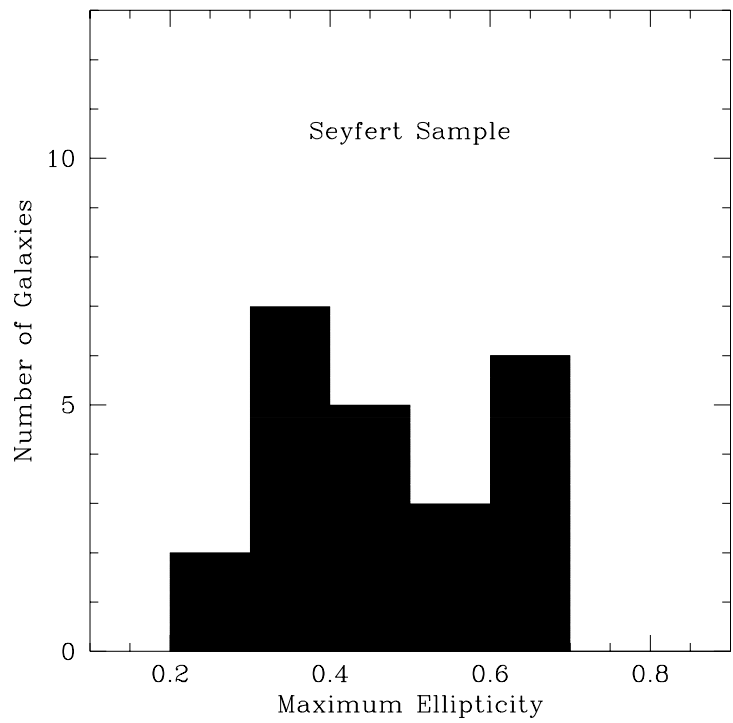
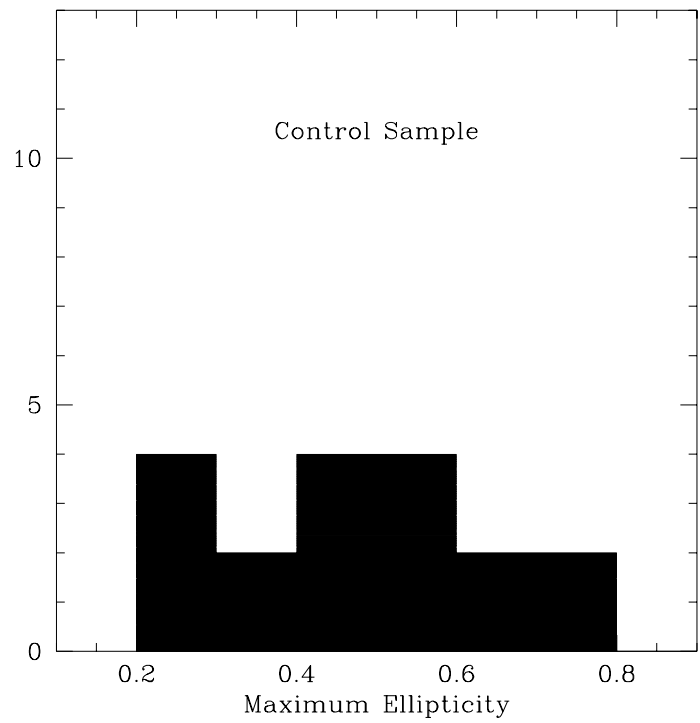
\end{document}